\documentclass[reprint, amsmath,amssymb, aps]{revtex4-1}

\usepackage{graphicx}
\usepackage{dcolumn}
\usepackage{bm}

\begin{document}

\preprint{APS/Ashish-Bhateja-2017}
\title{Rheology of dense granular flows in two dimensions: Comparison of fully two-dimensional flows to unidirectional shear flow}
\author{Ashish Bhateja}
\author{Devang V. Khakhar}
\email{khakhar@iitb.ac.in}
\affiliation{Department of Chemical Engineering, Indian Institute of Technology Bombay, Powai, Mumbai 400076, India}

\date{\today}
\begin{abstract}

This work utilizes soft-particle discrete element simulations to examine the rheology of steady two-dimensional granular flows with reference to a unidirectional shear flow, which has been extensively employed for validating the local visco-plastic model of Jop \textit{et al.} [Nature \textbf{441}, 727--730 (2006)]. The $\mu$-$I$ scaling proposed by Jop \textit{et al.} is found to be valid in both two-dimensional and unidirectional flows, as observed in previous studies, however, each flow type results in a different curve. Here $\mu$, ratio of the shear stress magnitude to the pressure, is the friction coefficient and $I$ is the dimensionless inertial number, which is proportional to the ratio of the magnitude of the rate of strain tensor, $\dot{\gamma}$, to the square root of the pressure. The friction coefficient is found not to scale in a simple way with the flow classification parameter $\psi$, which characterizes the local flow type. All the data collapse to a single curve using the scaling proposed by Zhang and Kamrin [Phys. Rev. Lett. \textbf{118}, 058001 (2017)], in which the scaled granular fluidity ($f=1/(\mu T)$, where $T \propto u/\dot{\gamma}$ and $u$ is the fluctuation velocity) is found to depend only on the solid fraction $\phi$. The data for variation of $\phi$ with inertial number $I$ collapse to a single curve for all the flows.
\end{abstract}

\maketitle
The development of theories for granular flows, in which the granular material is treated as a continuum, has a long history and many successes \cite{drucker1952, jenkins1983, mancini1987, campbell1990, lun1991, nedderman, degennes1999, mohan1999, midi2004, dacruz2005, jop2006, forterre2008, kamrin2012, henann2013, bouzid2015, jop2015, delannay2017}. The approach is particularly attractive for application to large systems, natural or industrial, which are comprised of billions of particles, making a particle scale analysis very expensive \cite{cleary2002,cleary2004}. A key element of such granular flow theories is a constitutive model to describe the rheology of the flowing material, and several models have been proposed based on different approaches and assumptions \cite{campbell1990, nedderman, campbell2006, forterre2008, jop2015, bouzid2015, delannay2017}. In the case of dense flows, a local visco-plastic model, proposed by Jop \textit{et al.} \cite{jop2006} has been validated for a number of systems, including for mixtures \cite{rognon2007, tripathi2011}, non-spherical particles \cite{mandal2016} and unsteady flows \cite{lacaze2009}.  Most of the systems used in the validation of the visco-plastic model correspond to unidirectional flows, an exception being the work of \citet{lacaze2009}, who showed its validity for the collapse of a cylindrical granular column, which is a three-dimensional unsteady flow. In addition, previous two-dimensional \cite{staron2012, staron2014} and three-dimensional \cite{kamrin2010} investigations also largely validate the visco-plastic model by comparing the velocity and stress fields from the discrete element simulations \cite{cundall79} with that of continuum predictions in a discharging silo. However, the model is known to break down in many instances, particularly, for slow creeping flows \cite{seguin2016}, in the presence of shear-localization near boundaries \cite{midi2004}, and in the case of thickness-dependent response of grains over an incline \cite{pouliquen2009,kamrin2015}. A more recent approach based on granular fluidity has shown promise to address several of the above issues \cite{kamrin2012,henann2013,henann2014,kamrin2015,kamrin2017}.
 
\begin{figure}
\begin{center}
\includegraphics[width=3.2in]{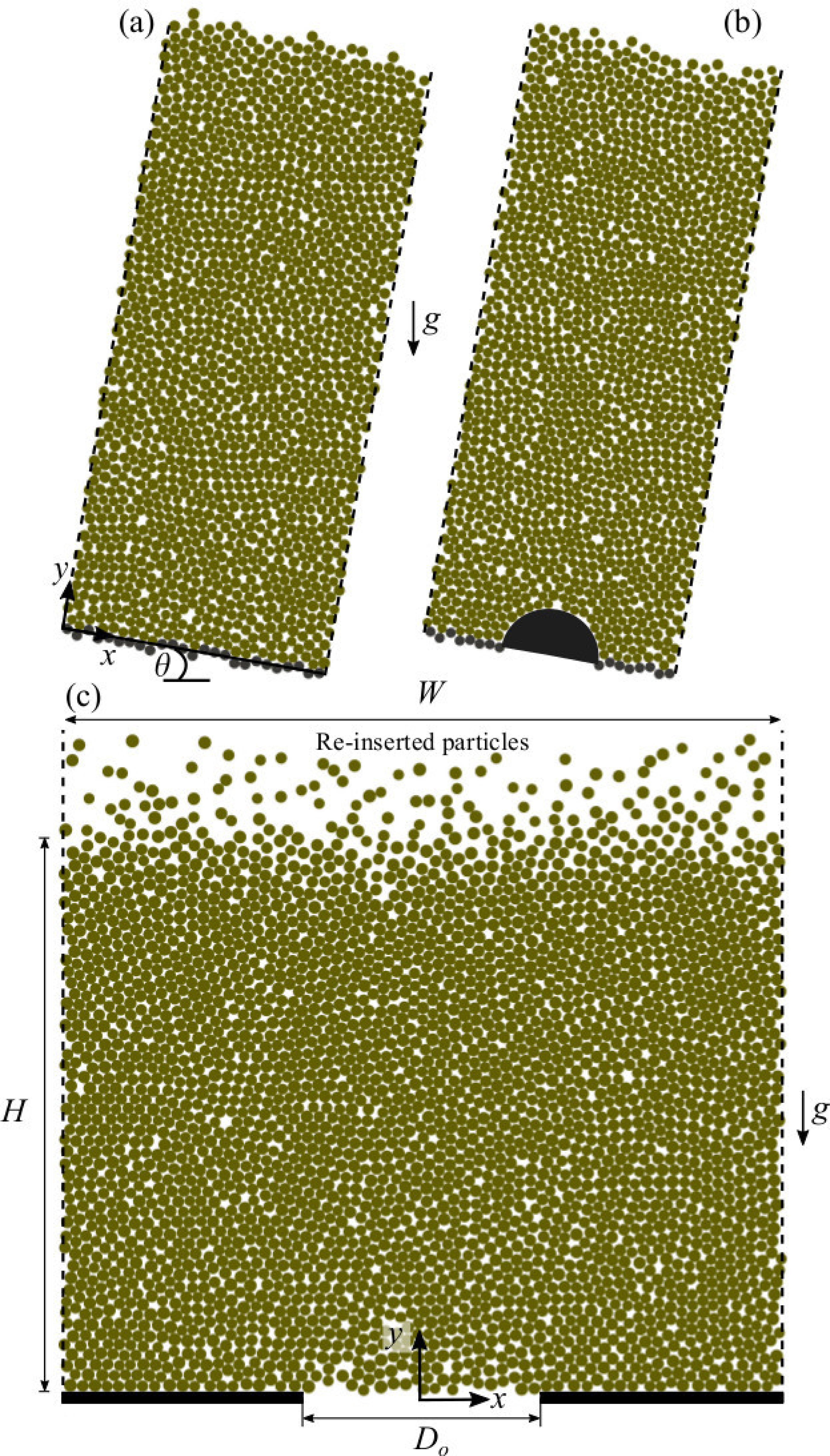}
\caption{\textit{Simulation snapshots}: (a) A classical unidirectional granular flow down a bumpy inclined surface. (b) Unidirectional flow on a bumpy inclined surface with an intruder. (c) A two-dimensional discharging silo. The coordinate axes and the direction of gravitational acceleration ($g$) are also displayed.}
\label{fig:snapshot}
\end{center}
\end{figure}

In this work, we characterize in detail the rheology of steady, two-dimensional granular flows in two different geometries (Fig. \ref{fig:snapshot}), with the flow over a rough inclined surface (Fig. \ref{fig:snapshot}(a))  chosen as a reference for comparison. The objective of the work is to understand the applicability of models developed for simple shear flows to fully two-dimensional flows using identical particles in different geometries of flow. The flow on a rough inclined surface with an intruder (Fig. \ref{fig:snapshot}(b)) represents a small deviation from a shear flow and the flow in a planar silo (Fig. \ref{fig:snapshot}(c)) is a fully two dimensional flow. Particles flowing out of the silo are uniformly reinserted at the top boundary of the silo at zero velocity, to maintain a constant height in the silo. Periodic boundary conditions are applied at the side boundaries  (dashed lines in Fig. \ref{fig:snapshot}) in all the cases. The data are used to evaluate the viscoplastic model \cite{jop2006} as well as the granular fluidity based model \cite{kamrin2017}.

Soft-particle discrete element computations \cite{cundall79} are utilized, considering grains as inelastic, frictional and non-cohesive disks of mean diameter $d$ with $\pm 10\%$ polydispersity so as to prevent ordering in the system. The mass density $\rho$ is kept the same for all grains. The interaction between particles is modelled as a linear-spring-dashpot force along with a Coulomb friction force \cite{cundall79,shafer1996,zhang1996,poschel}. The normal spring stiffness is $k_n = 10^6 \, mg/d$, and no tangential spring is employed, i.e., $k_t=0$. The restitution and friction coefficients for interaction between grains are $e_p = 0.9$ and $\mu_p = 0.4$, respectively. The same values are considered for grain-wall interactions. In the case of flow on a bumpy inclined surface, the length and the layer height along $x$ and $y$ directions are $25d$ and $56d$, respectively. The bumpy base, consisting of randomly placed immovable particles of size $d$, is inclined at an angle $\theta$ with the horizontal (see Fig.~\ref{fig:snapshot}(a)). We choose $\theta=13.5^{o}-19.5^{o}$, for which steady and fully-developed flow occurs. Moreover, the range of shear rates achieved is roughly the same as that obtained for the silo. In the case of the inclined surface flow with an intruder, the intruder radius is $5d$ and three angles $\theta=18^{o},19^{o}$ and $19.5^{o}$ are considered for this case. The silo width $W$ and the initial fill height $H$ of grains are $60d$ and $46d$, respectively. The orifice size $D_o$ is varied between $10d$ and $30d$ in steps of $5d$, with $D_o$ always being larger than $6d$ so as to prevent stoppage of flow due to arch formation \cite{mankoc2007,kondic2014}. The number of grains used in the inclined flow and inclined flow with intruder is $N=1500$, whereas $N=3000$ grains are employed in the silo flow.  

At steady state, the data are averaged over 20, 100 and 200 simulation runs for the inclined flow, inclined flow with an intruder and silo flow, respectively, with each simulation beginning with a different initial configuration. The averaging procedure follows the coarse-graining technique \cite{goldhirsch2010, weinhart2013, artoni2015}, employing a Heaviside step function with coarse-grained width $w$ equal to $d$. The stress tensor ($\bm{\sigma}$) is computed as given in Tripathi and Khakhar \cite{tripathi2010}. The quantities of interest do not change upon varying $w$ in the range $1d-5d$. In particular, we ensured independence of fluctuation velocity on $w$ following Artoni and Richard \cite{artoni2015}. 

\begin{figure}
\begin{center}
\includegraphics[width=3.2in]{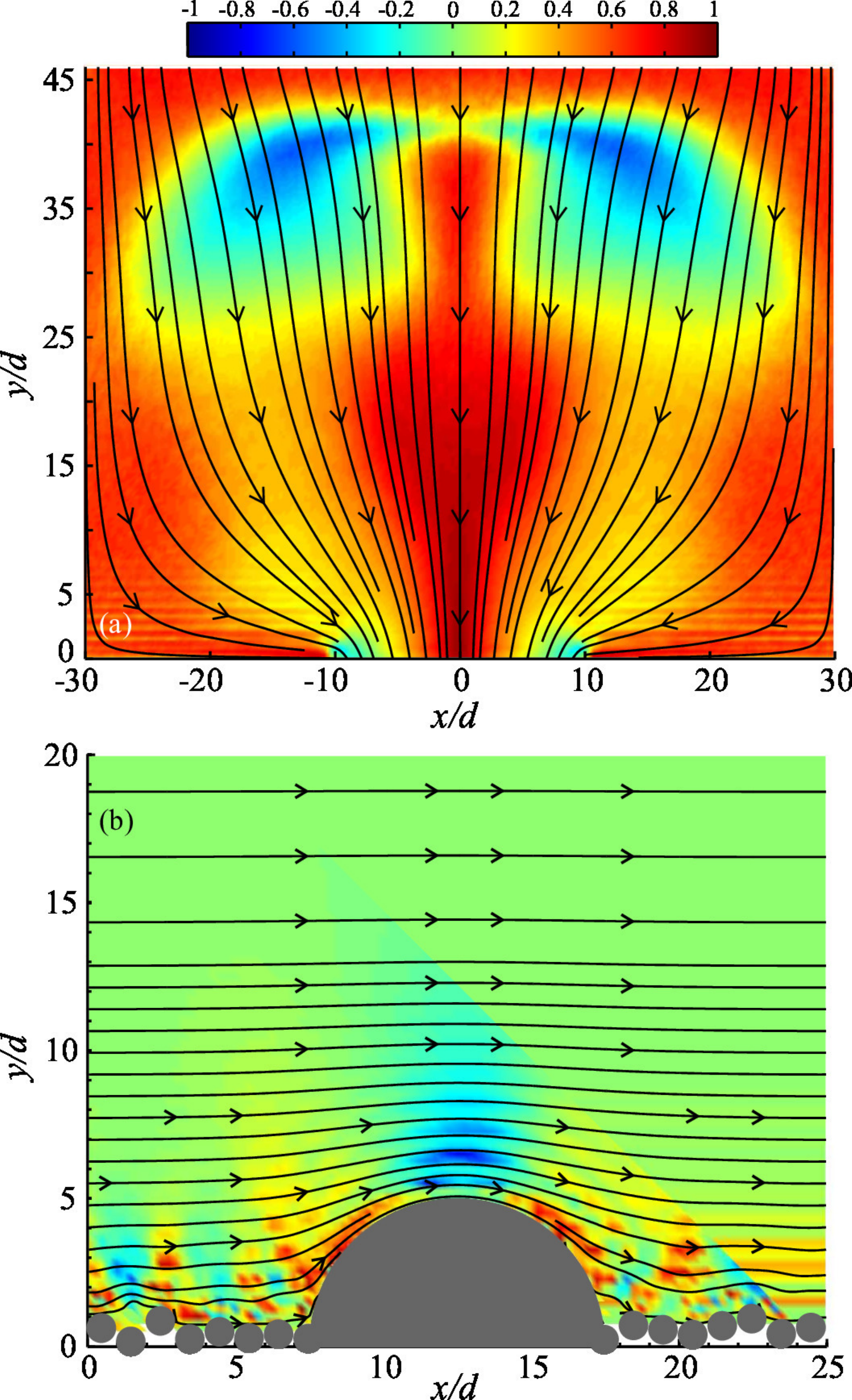}
\caption{Streamlines and spatial distribution of $\psi$ in case of (a) the discharging silo for $D_o = 20d$, and (b) the intruder-flow at 
$\theta=19.5^{o}$. Qualitatively similar distribution is obtained for other orifice sizes and inclinations for the silo and intruder-flow, respectively. The color scale for both plots is provided on top.}
\label{fig:psi}
\end{center}
\end{figure}

Fig.~\ref{fig:psi} shows the streamlines for the two-dimensional flows at steady state. In the inclined surface flow with an intruder, the streamlines near the free surface are straight lines as in the case of a uni-directional shear flow. However, near the base, the streamlines deviate from the shear flow case and a two-dimensional flow is obtained with compression of the streamlines above the intruder. The irregular streamlines at the base reflect the bumpy nature of the base. In the case of the silo, the streamlines converge into the orifice. No stagnant regions are seen because of the smooth base. Fig.~\ref{fig:psi} also shows the nature of the local flow, characterized in terms of a local flow parameter \cite{hudson2004,lee2007,wagner2016}, $\psi$, defined below. Two-dimensional isochoric flows may be linearized and transformed to the following form
\begin{equation}
v_{x^{'}}=\dot{\gamma}{y^{'}}, \qquad v_{y^{'}}=\psi\dot{\gamma}{x^{'}},
\end{equation}
where $(v_{x^{'}},v_{y^{'}})$ are the velocity components in the transformed coordinate directions ($x^{'},y^{'}$), $\dot{\gamma}$ is the shear rate and $\psi$ is a parameter which describes the nature of the flow: $\psi=0$ corresponds to shear flow, $\psi=1$ to pure extensional flow and $\psi=-1$ to solid body rotation. 

The flow parameter is the ratio of the eigenvalues of $\bm{Q}^T\bm{\cdot G}$, where $\bm{G}=[\nabla\bm{v}-(\nabla\bm{\cdot v})\bm{I}/2$] is the traceless velocity gradient tensor, $\bm{v}$ is the velocity vector, $\bm{I}$ is the unit tensor and $\bm{Q}=[0,-1;1,0]$ is the rotational matrix. This approach gives the same result as that of Hudson \textit{et al.} \cite{hudson2004} and Lee \textit{et al.} \cite{lee2007} for the calculation of the flow parameter ($\psi$). The flow parameter varies spatially in both two-dimensional flows (see Fig.~\ref{fig:psi}). In the inclined surface flow with an intruder, the flow is extensional ($\psi\approx1$)  just above the intruder but is close to a shear flow ($\psi=0$) in most of the region. The variation of $\psi$ is over a greater range in the silo flow. The flow is extensional ($\psi\approx1$) near the exit along the centreline and the magnitude of $\psi$ reduces with distance from the centerline. There are some regions in which $\psi<0$ indicating that the flow in these regions is more rotational than a shear flow. Thus, the two-dimensional flows chosen span a wide range of flow types. We note that, as shown in Sec.~A of Supplementary Material below, the scaled dilation rate, $\epsilon = (\nabla\bm{\cdot v})/\dot{\gamma}$, is small in both cases (less than 5\%) except in small regions of the flow, indicating that the flows are nearly isochoric, where $\dot{\gamma}=\sqrt{2\bm{D}:\bm{D}}$ with $\bm{D}=(\bm{G}+\bm{G}^T)/2$ being the rate of deformation tensor.

We consider the rheology of the above systems in terms of the model of \citet{jop2006} in which the material is assumed to be a Bingham fluid given by $\bm{\tau}=2\eta\bm{D}$ for $|\bm{\tau}| >\tau_s$, where $\bm{\tau}=\bm{\sigma}-P\bm{I}$ is the deviatoric stress tensor, $P=\mbox{tr}(\bm{\sigma})/2$ is the pressure, and $\eta$ is the viscosity, given in terms of the effective friction coefficient as $\eta=\mu P/\dot{\gamma}$. The friction coefficient is defined as $\mu=|\bm{\tau}|/P$, where $|\bm{\tau}|=(\bm{\tau:\tau}/2)^{1/2}$. We note that in the case of two-dimensional systems, the pressure and shear stress magnitude are related to the principal components of the stress tensor ($\sigma_1,\sigma_2$) as $P=(\sigma_1+\sigma_2)/2$ and $|\bm{\tau}|=(\sigma_1-\sigma_2)/2$, so that $\mu=(\sigma_1-\sigma_2)/(\sigma_1+\sigma_2)$. In the limit of no flow, we have $\mu=\mu_s=\tau_s/P$ and on applying the Mohr-Coulomb failure criterion in this case, we have $\mu_s=\sin(\beta_s)$, where $\beta_s$ is the angle of internal friction. An assumption of the model of \citet{jop2006} is that the friction coefficient ($\mu$) depends only on the \textit{inertial number} defined as $I=\dot{\gamma}d/\sqrt{P/\rho}$, and the following phenomenological relationship relating the two was proposed
\begin{equation}
\mu(I) = \mu_s + \frac{\mu_m - \mu_s}{(1+I_0/I)}
\label{eqn:mulaw}
\end{equation}
where $\mu_s, \mu_m$ and $I_0$ are fitting parameters. The dilatancy of the flow is also assumed to depend only on the inertial number and the solid fraction is given by a power-law expression \cite{hatano2007}
\begin{equation}
\phi = \phi_{m} -  b~I^{n},
\label{eqn:phi}
\end{equation}
where $\phi_{m}$, $b$ and $n$ are fitting parameters.

\begin{figure}
\begin{center}
\includegraphics[width=3.2in]{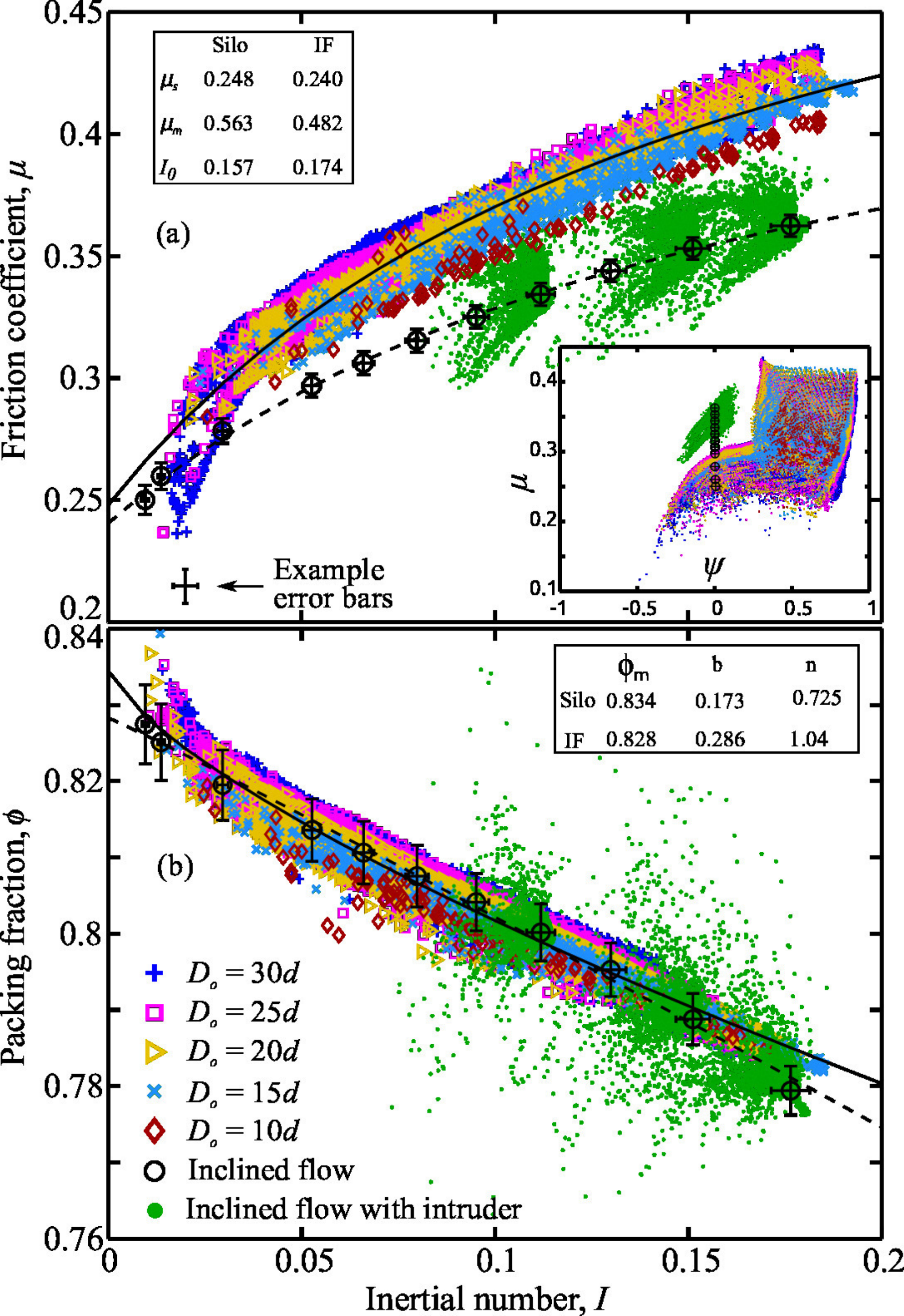}
\caption{(a) Variation of $\mu$ with $I$. The example error bars for $\mu$ and $I$ at their highest values are displayed on the bottom left corner of the plot for silo. The solid (silo) and dashed (inclined flow) lines are fits of Eq. \ref{eqn:mulaw}. \textit{Inset}: Variation of $\mu$ with $\psi$. Points are used instead of symbols for ease of presentation, keeping colors the same as used for symbols for corresponding $D_o$. Relative error constraint is imposed only on $\mu$, not on $\psi$. Note that (average) $\psi=0$ for inclined flow (simple shear). (b) Variation of $\phi$ with $I$. The solid (silo) and dashed (inclined flow) lines are fits of Eq. \ref{eqn:phi}. The respective fitting parameters for $\mu$ and $\phi$ are specified in the rectangular boxes.}
\label{fig:muphi}
\end{center}
\end{figure}

\begin{figure}
\begin{center}
\includegraphics[width=3.2in]{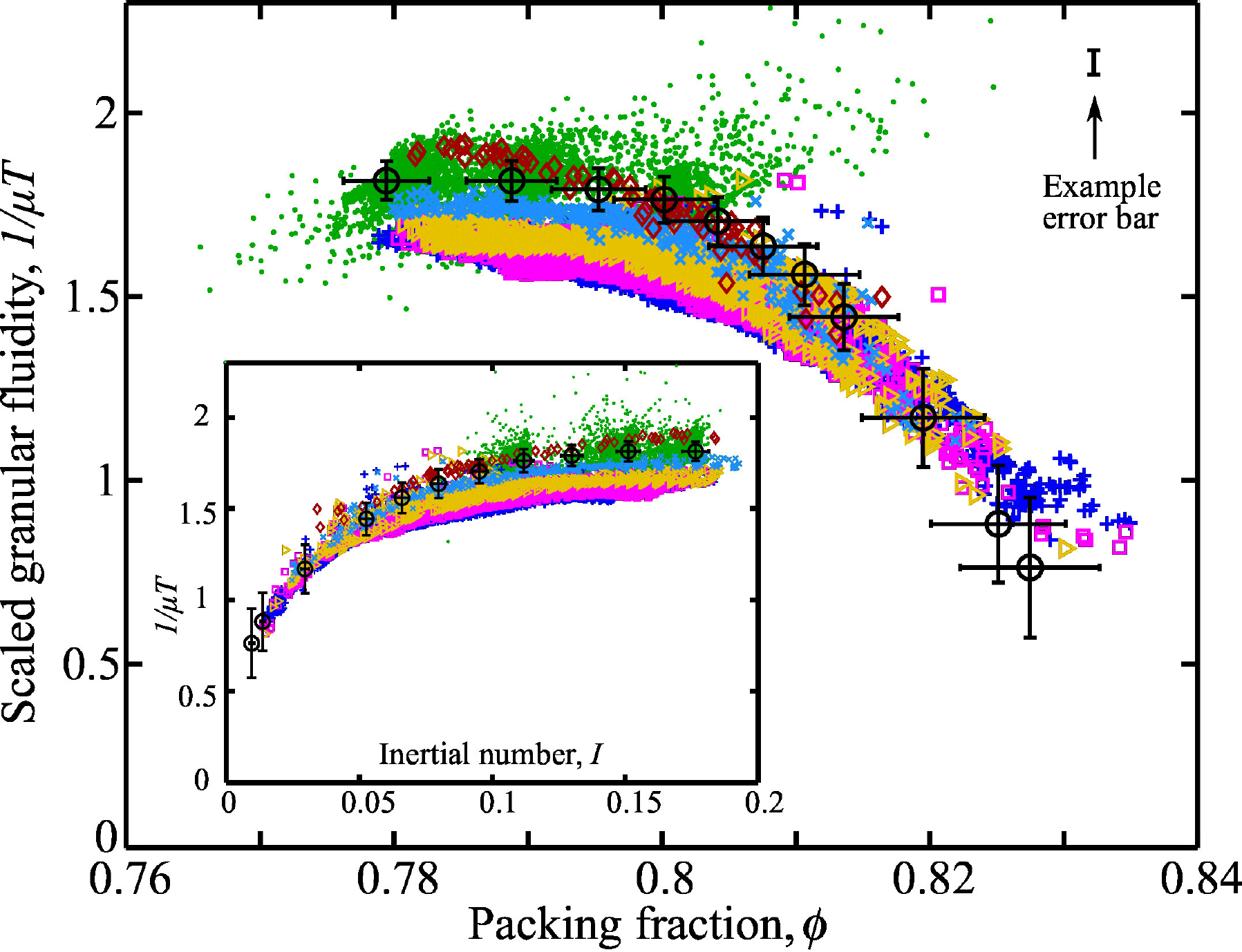}
\caption{Scaled granular fluidity $1/\mu T$ versus $\phi$. The example error bar for $1/\mu T$ at its highest value is displayed on top right corner for silo. Inset shows variation of $1/\mu T$ with $I$. The legend is provided in Fig. \ref{fig:muphi}.}
\label{fig:muT}
\end{center}
\end{figure}

We found that the stress tensor was symmetric, ruling out Cosserat effects \cite{mohan1999}, and that the principal directions of the stress tensor ($\bm{\sigma}$) and rate of deformation tensor ($\bm{D}$) were nearly coaxial (see Sec.~B of Supplementary Material below), as also reported by \citet{rycroft2009}, indicating the validity of the tensor form of the constitutive equation proposed by \citet{jop2006}. Fig.~\ref{fig:muphi}(a) shows the variation of the effective friction coefficient ($\mu$) with the inertial number ($I$) for all three systems studied and Fig.~\ref{fig:muphi}(b) gives variation of the packing fraction ($\phi$) with $I$.  Data points lying within $5d$ and $10d$ of the boundaries and free surface, respectively, and those points where relative error of any variable is more than 1.5\% are not shown in Figs.~\ref{fig:muphi} and \ref{fig:muT}. The data for the inclined surface flow follow previously reported results \cite{forterre2008,jop2015}. The silo data appear to collapse quite well for different orifice sizes, as also reported by Lacaze \& Kerswell \cite{lacaze2009} and \citet{cortet2009} for inhomogeneous flow configurations other than the silo, but do not fall on the inclined flow data.  The data for the inclined surface flow with an intruder lie between the two data sets. Equation~(\ref{eqn:mulaw}) describes each data set quite well, and the fitted values of the parameters are given in the legend. The inset in Fig.~\ref{fig:muphi}(a) shows the variation of the friction coefficient ($\mu$) with the flow parameter ($\psi$). There does not appear to be any correlation indicating that a modification of Eq.~(\ref{eqn:mulaw}) incorporating the flow parameter may not be a useful approach. In contrast, we obtain a very good collapse of data for the solid fraction ($\phi$) variation with the inertial number ($I$) for all the three systems considered and Eq.~(\ref{eqn:phi}) fits very well (Fig.~\ref{fig:muphi}(b)). The above results indicate that an additional parameter may be needed to describe the rheology of two-dimensional flows.

We next analyze the rheology in terms of the scaled granular fluidity ($f$), which for steady granular flows is found to depend  only on the solid fraction ($\phi$) \cite{kamrin2017}. In the present notation,  the scaled granular fluidity is given by $f=1/(\mu T)$, where $T = u/\dot{\gamma}d$ and $u$ is the root mean square (r.m.s.) fluctuation velocity. Fig.~\ref{fig:muT} shows the variation of $1/\mu T$ with $\phi$ for all three flow configurations. The data collapse to a single curve reasonably well and the variation obtained is similar to what is obtained by Zhang and Kamrin \cite{kamrin2017}, i.e., $f$ is nearly constant at low $\phi$ and decreases in close to linear fashion at large packing densities. Importantly, the range of $f$ obtained in the current two-dimensional systems is close to what is obtained in \cite{kamrin2017} for three dimensional flows. A minor departure is seen for smaller orifice sizes at low $\phi$. Given the close correlation between the solid fraction and the inertial number, we replot the data in Fig.~\ref{fig:muT} in terms of the inertial number in the inset. The collapse of the data for $I$ appears to be better in this case. 

We investigated in detail, by means of numerical simulations, the rheology of steady, dense granular flows in three different planar geometries of increasing complexity. The local nature of the flows, as determined by the flow parameter ($\psi$), is shown to vary considerably, spanning the range from pure rotational flow to pure extensional flow. The data are analyzed in terms of the model of \citet{jop2006} and the $\mu$--$I$ scaling is found to be valid for each geometry, but the data for the three geometries do not collapse to a single curve. Thus the scaling does not extend to two-dimensional flows. Analyzing the data in terms of the scaled granular fluidity yields a better collapse of the data to a single curve. This finding indicates the importance of including velocity fluctuations in the constitutive relation. The solid fraction scales with inertial number quite well and all the data collapse to a single curve, with $\phi(I)$ for all the geometries considered.

We thank Neeraj Kumbhakarna for providing access to his computational cluster for running simulations presented in this paper. Ashish Bhateja is grateful to Ken Kamrin for a useful discussion. Financial support of IIT Bombay and SERB, India (Grant No. SR/S2/JCB-34/2010) is gratefully acknowledged.

\section*{Supplementary Material}
\subsection{Scaled dilation rate}

\begin{figure}[ht!]
\begin{center}
\includegraphics[scale=0.45]{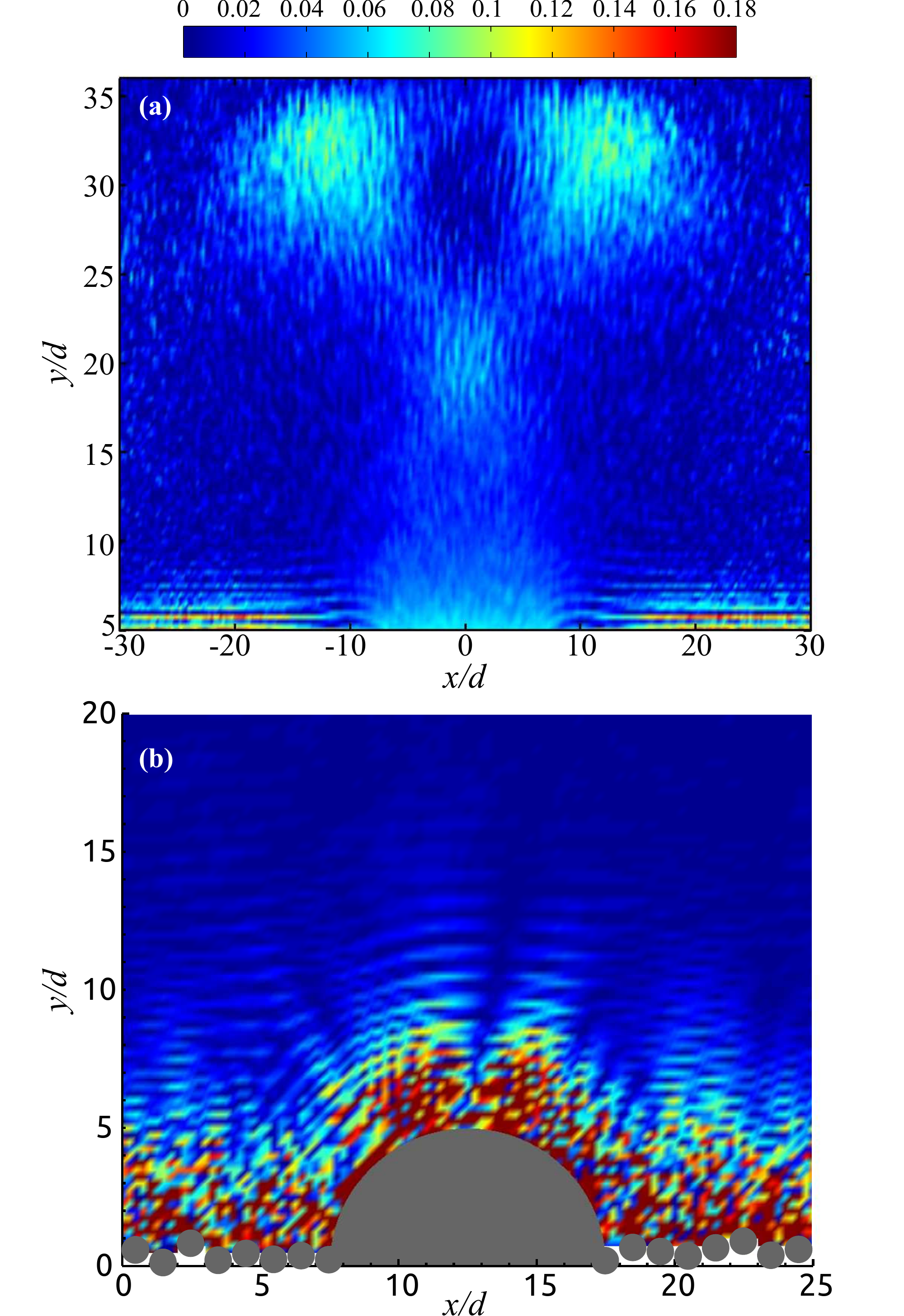}
\caption{Spatial distribution of scaled dilation rate $\epsilon=\nabla \cdot \bm{v}/\dot{\gamma}$ in case of (a) the discharging silo for $D_o = 20d$  and (b) the inclined flow with intruder for $\theta=19.5^o$. Qualitatively similar distribution is obtained for other orifice sizes and inclinations as well in the case of silo and inclined flows, respectively. The color scale for both plots is provided on top.}
\label{fig:drate}
\end{center}
\end{figure}

Figure~\ref{fig:drate} displays spatial distribution of scaled dilation rate $\epsilon=\nabla \cdot \bm{v}/\dot{\gamma}$ for the silo and inclined flow with intruder. The distribution is shown in the case of silo for domain excluding the points lying within $5d$ and $10d$ of the base and free surface, respectively. This region is that we consider for plotting $\mu$, $\phi$ and $1/\mu T$ with inertial number $I$ in Figs.~ 3 and 4 of the main text. The spatial distribution for the inclined flow is provided in the same vertical range as what is given for $\psi$ in the main text in Fig.~2, i.e., from the base to $y=20d$. We note that $\epsilon$ is largely below $5\%$ in both flows except for a few regions lying close to the base in the silo and in proximity to the intruder in the case of inclined flow. This demonstrates that the flow is nearly incompressible in the region that we consider for rheological measurements.
\subsection{Principal direction}

\begin{figure}[ht!]
\begin{center}
\includegraphics[scale=0.4]{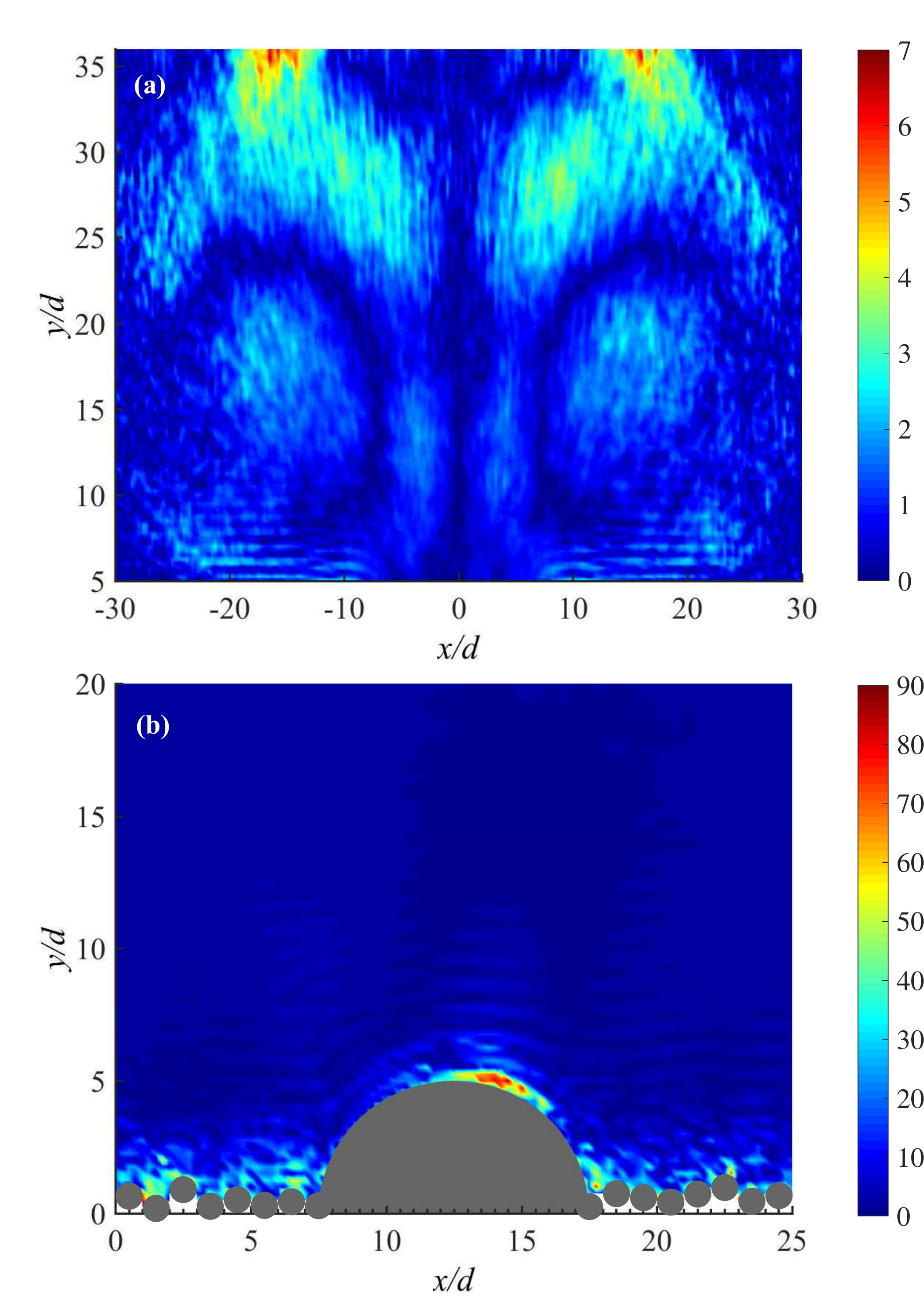}
\caption{Spatial distribution of the angle ($\alpha$) between the principal directions of $\bm{\sigma}$ and $\bm{D}$ in case of (a) the silo for $D_o=20d$ and (b) the inclined flow with intruder for $\theta=19.5^o$. Color scales for both plots are provided separately on their right side and the values on the color scales are given in degrees.}
\label{fig:pdir}
\end{center}
\end{figure}

Figure \ref{fig:pdir} shows spatial distribution of the angle ($\alpha$) between a principal direction of the stress tensor $\bm{\sigma}$ with that of the rate of deformation tensor $\bm{D}$, in the case of discharging silo for $D_o=20d$ and inclined flow with intruder for $\theta=19.5^{o}$. Again, as mentioned in the previous section, the data are considered for the domain excluding the points lying within $5d$ and $10d$ of the base and free surface, respectively. As displayed in Fig.~\ref{fig:pdir}, the principal directions of $\bm{\sigma}$ and $\bm{D}$ nearly aligns with each other as $\alpha$ is mostly below $5^{o}$, barring a small fraction of the domain at the top in the case of silo (see Fig.~\ref{fig:pdir}(a)) and adjacent to the intruder in the case of inclined flow (see Fig.~\ref{fig:pdir}(b)). This shows largely the existence of coaxiality in the most of the flow region under investigation.

\providecommand{\noopsort}[1]{}\providecommand{\singleletter}[1]{#1}%
%

\end{document}